\newcommand{\PRL}[3]{Phys.\ Rev.\ Lett.\ {\bf #1},\ #2 (#3)}
\newcommand{\RMP}[3]{Rev.\ Mod.\ Phys.\ {\bf #1},\ #2 (#3)}

\newcommand{\NAT}[3]{Nature (London)\ {\bf #1},\ #2 (#3)}

\newcommand{\PRA}[3]{Phys.\ Rev.\ A\ {\bf #1},\ #2 (#3)}

\newcommand{\QO}[3]{Quantum\ Opt.\ {\bf #1},\ #2 (#3)}
\newcommand{\FOP}[3]{Fortschr.\ Phys.\ {\bf #1},\ #2 (#3)}
\documentclass[aps,showpacs,showkeys,amsmath,amsfonts,amssymb]{revtex4}
\usepackage{graphicx}

\begin{document}
\title{A proposal to generate entangled compass states with sub-Planck
structure}
\author{Sayan Choudhury}
\email{sayan.cho@gmail.com}

\author{Prasanta K. Panigrahi}
\email{prasanta@prl.res.in}
\affiliation{Indian Institute of Science Education and Research Kolkata,
Mohanpur Campus, BCKV Campus Main Office, Mohanpur - 741252, India}

\begin{abstract}
We illustrate a procedure to generate a bipartite, entangled compass state,
which shows sub-Planck structure. The proposed method uses the interaction of a
standing wave laser field, with two, two-level atoms and relies on the ability of
this system to choose certain mesoscopic bipartite states to couple with the
internal degrees of freedom. An appropriate
measurement on the internal degrees of freedom then leads to the entangled state,
which shows sub-Planck structures, desired for quantum metrology.
\end{abstract}
\maketitle
\section{introduction}
In recent times, macroscopic quantum states with non-local superpositions have
attracted considerable attention in the context of quantum metrology. It was
demonstrated by Zurek that, the single particle compass state shows sub-Planck
sensitivity, which makes it useful for carrying out Heisenberg-limited
measurements \cite{Zurek}. The sub-Planck structure owes its origin to
interference in phase space \cite{Wheeler}. A criterion for distinguishing quantum states, which do not have a classical counterpart, from those which have one, has been studied in Refs. \cite{Vogel, Richter}. Therein structures in phase space that are narrower than their ground state counterparts have also been demonstrated in the context of an oscillator. A classical system possessing
sub-Fourier sensitivity has been experimentally realized \cite{Praxmeyer}.
Sub-Planck analogs have been identified in chronocyclic phase space \cite{Austin}. The existence of such structures has been found in the P{\"o}schl-Teller potential
\cite{Utpal}. The effect of decoherence on sub-Planck structures have been studied in Ref. \cite{Suranjana}. In the quantum scenario, a number of proposals have been advanced
for generating single particle cat and generalized states showing the above feature
\cite{GSA,Toscano,Ghosh,Milburn}. Agarwal and Pathak suggested a scheme for generating
the compass state in a cavity QED scenario, using a Rydberg atom coupled to a
single-mode high Q-cavity via the Jaynes-Cummings evolution, followed by joint
detection of two atoms in a particular state \cite{GSA}. The cat state, which
shows partial sensitivity for Heisenberg-limited measurements, can be generated
by non-linear optical processes \cite{Tanas, Tara} and by quantum-non-demolition
measurements of the photon number in cavity QED \cite{Brune} and the vibrational
quantum number of a trapped atom \cite{Schneider}. \\

It has been recently found that an appropriate entangled bipartite system can
help carry out Heisenberg-limited measurements with better sensitivity
\cite{panigrahi}. Furthermore, it may be more robust to decoherence. Since, the
compass state is a superposition of four constituent coherent states, it is
strongly prone to environmental effects \cite{GSA}. It is worth mentioning that
a number of schemes have been implemented for entangling macroscopic quantum
states, for quantum information processing \cite{Ketterle,Ketterle1}.
Entanglement of a mesoscopic
field with a Rydberg atom has been achieved experimentally in cavity QED
\cite{Haraoche}. Light mediated distribution of motional state entanglement has been proposed,
 in the context of harmonically trapped atoms \cite{Parkins}. Entangled states have been considered for measuring weak effects arising out of atomic parity violation \cite{Manas}.\\
In this paper, we make use of mesoscopic quantum states to demonstrate a procedure to generate general
entangled bipartite states having
sub-Planck sensitivity in both co-ordinate and momentum spaces. We
couple two identical neutral two-level atoms in a deep double well potential to
a standing wave laser field, with each well containing a single atom. Instead of
a double well, two independent harmonic traps or an optical double lattice which
has already been experimentally realized \cite{Ellman} can also be used. 
 This type of systems
have been proposed for carrying out quantum logic operations in an optical
lattice \cite{Hema} and for atoms on a chip \cite{Brennen,Jo}. In our scheme, the first atom
couples
to the laser field via the position of its center-of-mass, while the second one
interacts with the field after a suitable time-delay, amounting to a coupling
through the
velocity of its center-of-mass. This scheme can be implemented in cold atoms, such that the system is in its ground state. Otherwise, a sideband cooling may be effected for the same purpose. This is followed by joint detection of
the atoms in definite internal states. The depth of each well and distance between
the two
wells can be adjusted in order to control tunneling effects. In the following
section, we explicate the necessary steps involved in generating these compass
states and study their sub-Planck structures in phase space in section III. We
then  conclude with discussions regarding further investigation.

\section{Generation of bipartite entangled states with sub-Planck sensitivity}
It has been proposed recently that certain continuous variable bipartite
entangled states reveal sub-Planck structure {\cite{panigrahi}. These entangled
states contain cat-type constituent states and offer better sensitivity for
Heisenberg limited measurements than the original compass state. \\
The proposed bipartite compass state is of the type,
\begin{eqnarray}
|\psi\rangle_c&=&
a(|\alpha\rangle_{1}|i\alpha\rangle_{2}+|i\alpha\rangle_{1}|\alpha\rangle_{2})+
b(|-\alpha\rangle_{1}|-i\alpha\rangle_{2}+|-i\alpha\rangle_{1}|-\alpha\rangle_{2
})+\nonumber \\ & &
c(|\alpha\rangle_{1}|-i\alpha\rangle_{2}+|i\alpha\rangle_{1}|-\alpha\rangle_{2}
)+d(|-\alpha\rangle_{1}|i\alpha\rangle_{2}+|-i\alpha\rangle_{1}|\alpha\rangle_{2
}) ,
\end{eqnarray}
where $|\alpha\rangle$ is the coherent state:
\begin{equation}
\hat a |\alpha\rangle= \alpha|\alpha\rangle.
\end{equation}

As will be explicitly shown in the subsequent section, in order to produce
sub-Planck structures in the phase space, at least one of the tuples (a,b) or
(c,d) must have both the co-efficients non-zero.  \\

For a physical realization of the above states, we consider two identical
two-level
atoms in a deep double well potential, coupled to a standing wave laser field.
The excited and ground levels $|e\rangle$ and $|g\rangle$ are the internal
degrees of freedom of each atom. We assume that the probability of occupation of
the excited states of the double well is very small. Hence, to start with the wells containing
one atom each can be effectively treated as two independent harmonic traps. In the following, we assume that spontaneous emission from the upper internal level of the atoms can be neglected and furthermore decoherence of center-of-mass motion will also be ignored.\\
For a trapped atom in a standing wave laser field in the Lamb-Dicke regime, 
where the wavelength of the classical field is long compared to the extent of the
confining trap, the general time evolution is governed by \cite{Knight}:
\begin{equation}
\hat H=\hat H_{s}+\hat H_{\epsilon}+\hat H_{int}.
\end{equation}
Here,
\begin{equation}
H_{s}=\frac{\Delta}{2} \hat \sigma_{z},
\end{equation}
$\Delta$ being the atom-laser detuning parameter, which can be adjusted to zero.
 The Hamiltonian describing the motional degrees of freedom of the trapped atoms
undergoing harmonic motion is given by:
\begin{equation}
\hat H_{\epsilon} =\sum_{i=1} ^{2} \left( \hat p_{i}^{2}/2m + \frac{1}{2} m
\omega^{2}
\hat x_{i}^{2} \right).
\end{equation}
$H_{int}$ describes the coupling of the laser with the atoms; for the $i^{th}$ atom, this can be represented by \cite{Knight}:
\begin{equation}
\hat H_{int}=\lambda [cos(\phi) - \eta sin(\phi)]\hat \sigma_{x} ^{i} \otimes c 
\hat x_{i}.
\end{equation}
Here, $\hat \sigma_{x}=|g\rangle\langle e|+|e\rangle\langle g|$; $\hat x_{i}$
denotes the position of the center of mass of the $i^{th}$ atom and the relative
position of the trap in the field is determined by the parameter $\phi$. For
convenience of illustration, the position of the trap is tuned such that
$\phi=\frac{3 \pi}{2}$, whence the Hamiltonian takes the suggestive
form:
\begin{equation}
\hat H_{\epsilon} ^{i}= \frac{\hat p_{i}^{2}}{2m}+ \frac{1}{2} m \omega^{2}
[[\hat
x_{i}+\frac{c\sigma_{x} ^{i}}{m \omega^{2}}]^{2} ] -\frac{c^{2}}{m \omega^{2}},
\end{equation}
with $c=\lambda \eta $. As is evident, the present case is similar to
the spin-boson system, where a single
two-level system interacts with a large reservoir of bosonic field modes
\cite{Leggett}.\\
The two degenerate lowest energy eigenvectors are:
\begin{eqnarray}
|\zeta_{+}\rangle &=&|+\rangle_{i}|\alpha\rangle,\nonumber\\ 
\textrm {and\,\,}
 |{\zeta}_{-}\rangle &=&|-{\rangle}_{i}|-\alpha\rangle;
\end{eqnarray}
with,
\begin{eqnarray}
|+\rangle &=&\frac{1}{\sqrt 2}[|g\rangle + |e\rangle],\nonumber\\
\textrm{and \,\,}
|-\rangle &=&\frac{1}{\sqrt 2}[|g\rangle - |e\rangle].
\end{eqnarray}
In the above,
\begin{equation}
|\alpha\rangle=e^{-i\alpha \hat p/\hbar}|E_{0}\rangle.
\end{equation}
Here, $|E_{0}\rangle$ refers to the ground state of the harmonic oscillator and
$\alpha=\frac{c}{m\omega^{2}}$.

Through an unitary transformation \cite{Ford}, 

\begin{equation}
 U=exp[-i \frac{\pi}{4} \hat a^{\dagger}_{j} \hat a_{j}],
\end{equation}
the Hamiltonian in Eq.(7) can be equivalently represented by another one, where the atom couples to the laser field through the 
momentum of its center-of-mass. The above unitary operator corresponds to a free evolution for a period 
$\frac{\pi}{4}$, physically amounting to a delayed coupling between the
$j^{th}$atom and the standing wave laser field. It is assumed that the distance
of the second well can be adjusted for the free evolution of the atom located in the same. The
unitarily equivalent interaction Hamiltonian,
describing the above coupling between the laser field and the $j^{th}$ atom, is given by:
\begin{equation}
\hat H_{int}= \hat \sigma_{x} ^{j} \otimes d  \hat p_{j} ,
\end{equation}
with
\begin{equation}
d=\frac{c}{m \omega}.
\end{equation}
The two degenerate lowest energy eigenvectors for this system are,
\begin{eqnarray}
|\eta_{+}\rangle&=&|+\rangle_{j}|i\alpha\rangle. \\
\textrm{and\,\,}
|\eta_{-}\rangle&=&|-\rangle_{j}|-i\alpha\rangle.
\end{eqnarray}

The first atom
couples to
the laser field via the position of its center-of-mass, while the other atom
interacts with the field after a time-delay, amounting to coupling through the
velocity of its center-of-mass. The interaction Hamiltonian couples $|\alpha\rangle$ or
$|i\alpha\rangle$ to the $|+\rangle$ state depending on whether the coupling
occurs via the position or momentum respectively. Similarly, $|-\alpha\rangle$ or
$|-i\alpha\rangle$ gets coupled with the $|-\rangle$ state. Since, this scheme can be implemented in cold atoms, we assume that the system is in its ground state. Otherwise, a 
sideband cooling can be performed for the same purpose. Initially the wells are well separated so that
the probability of tunneling is very low; after both the traps have been
subjected to the standing wave laser field, the distance between the wells can
be adjusted, so as to allow tunneling. For identical bosonic atoms,
this
would select out the symmetric state as the ground state of the system, which
modulo normalization, can be written in the form:
\begin{eqnarray}
|\psi\rangle &=&
a|+\rangle|+\rangle(|\alpha\rangle_{1}|i\alpha\rangle_{2}+|i\alpha\rangle_{1}
|\alpha\rangle_{2})+
b|-\rangle|-\rangle(|-\alpha\rangle_{1}|-i\alpha\rangle_{2}+|-i\alpha\rangle_{1}
|-\alpha\rangle_{2})+\nonumber \\ & &
c(|-\rangle_{1}|+\rangle_{2}(|\alpha\rangle_{1}|-i\alpha\rangle_{2}+|+\rangle_{1
}|-\rangle_{2}|-i\alpha\rangle_{1}|\alpha\rangle_{2})+d(|+\rangle_{1}|-\rangle_{
2}(|-\alpha\rangle_{1}|i\alpha\rangle_{2}+|-\rangle_{1}|+\rangle_{2}
|i\alpha\rangle_{1}|-\alpha\rangle_{2}).
\end{eqnarray}

At this stage, a joint detection of the internal states of both the atoms can produce four different
types of entangled bipartite states of the desired sub-Planck sensitivity. This
is shown in the table below. For illustration, we analyze the state, which is
obtained when both the atoms are in the excited state.  \\
\begin{table}[h]
\caption{\label{tab9} Outcome and the obtained state (unnormalized)}
\begin{tabular}{|c|c|}
\hline {\bf Outcome of the Measurement } & {\bf State obtained}\\
\hline
$|ee\rangle$&$
a(|\alpha\rangle_{1}|i\alpha\rangle_{2}+|i\alpha\rangle_{1}|\alpha\rangle_{2})+
b(|-\alpha\rangle_{1}|-i\alpha\rangle_{2}+|-i\alpha\rangle_{1}|-\alpha\rangle_{2
})+$ \\ 
$ $&$
c(|\alpha\rangle_{1}|-i\alpha\rangle_{2}+|i\alpha\rangle_{1}|-\alpha\rangle_{2}
)+d(|-\alpha\rangle_{1}|i\alpha\rangle_{2}+|-i\alpha\rangle_{1}|\alpha\rangle_{2
})$\\
$ $&$ $\\
$|gg\rangle$&$
a(|\alpha\rangle_{1}|i\alpha\rangle_{2}+|i\alpha\rangle_{1}|\alpha\rangle_{2})+
b(|-\alpha\rangle_{1}|-i\alpha\rangle_{2}+|-i\alpha\rangle_{1}|-\alpha\rangle_{2
})-$\\ 
$ $&$
c(|\alpha\rangle_{1}|-i\alpha\rangle_{2}+|i\alpha\rangle_{1}|-\alpha\rangle_{2}
)-d(|-\alpha\rangle_{1}|i\alpha\rangle_{2}+|-i\alpha\rangle_{1}|\alpha\rangle_{2
})$\\
$ $&$ $\\
$\frac{1}{\sqrt2}(|ge\rangle+|eg\rangle)$&$
a(|\alpha\rangle_{1}|i\alpha\rangle_{2}+|i\alpha\rangle_{1}|\alpha\rangle_{2})+
b(|-\alpha\rangle_{1}|-i\alpha\rangle_{2}+|-i\alpha\rangle_{1}|-\alpha\rangle_{2
})$\\
$ $&$ $\\
$\frac{1}{\sqrt2}(|ge\rangle-|eg\rangle)$&$
c(|\alpha\rangle_{1}|-i\alpha\rangle_{2}+|i\alpha\rangle_{1}|-\alpha\rangle_{2}
)-d(|-\alpha\rangle_{1}|i\alpha\rangle_{2}+|-i\alpha\rangle_{1}|\alpha\rangle_{2
})$\\
\hline
\end{tabular}
\end{table} \\

\section{sub-Planck sensitivity of the compass state}
We now proceed to study the phase space structures of $|\psi_{c}\rangle$ for
ensuring sub-Planck sensitivity.
For calculational simplicity and in order to establish correspondence with
earlier results \cite{panigrahi}, we rewrite $\psi_{c}$ in the form:
\begin{eqnarray}
\vert\psi\rangle_c&=&A[\frac{1}{\sqrt 2}(\vert \pm\alpha^{+} \rangle_1\vert \pm i\alpha^{+}
\rangle_2
+\vert \pm i\alpha^{+} \rangle_1\vert \pm\alpha^{+} \rangle_2)]+B[
\frac{1}{\sqrt 2}(\vert \pm\alpha^{-} \rangle_1\vert \pm i\alpha^{-}+\vert \pm
i\alpha^{-} \rangle_1\vert \pm\alpha^{-})]+ \nonumber \\ & & C[\frac{1}{\sqrt
2}(\vert \pm\alpha^{+} \rangle_1\vert \pm i\alpha^{-}\rangle_{2}+\vert \pm
i\alpha^{-} \rangle_1\vert \pm\alpha^{+}\rangle_{2})]+D[\frac{1}{\sqrt 2}(\vert
\pm\alpha^{-} \rangle_1\vert \pm i\alpha^{+}\rangle_{2}+\vert \pm i\alpha^{+}
\rangle_1\vert \pm\alpha^{-}\rangle_{2})]
\end{eqnarray}
where, {A, B, C, D} and {a, b, c, d} are linearly related.\\
Here,

\begin{equation}
\vert \pm\alpha^{+} \rangle=\frac{1}{\sqrt{2}}(\vert \alpha\rangle
+\vert-\alpha\rangle)
\end{equation}
and
\begin{equation}
\vert \pm\alpha^{-} \rangle=\frac{1}{\sqrt{2}}(\vert \alpha\rangle
-\vert-\alpha\rangle).
\end{equation}
These states do not generically satisfy separability criterion  based
on the variance approach \cite{Simon,Duan} and can be represented by localized
Gaussian states. In the
coordinate basis, $\langle x\vert \pm\alpha^{\pm}\rangle = \psi^{\pm}(x)$ 
and $\langle x\vert \pm i\alpha^{\pm}\rangle =
\varphi^{\pm}(x)$, where 
\begin{equation}
\psi^{\pm}(x)=\frac{e^{-(x+x_{0})^{2}/2\delta^{2}}\pm
e^{-(x-x_{0})^{2}/2\delta^{2}}}
{\sqrt{2}\pi^{1/4}\delta^{1/2}\left[ 1\pm
e^{-x_{0}^{2}/\delta^{2}}\right]^{1/2}}
\end{equation}
and
\begin{equation}
\varphi^{\pm}(x)=\frac{e^{-x^{2}/2 \delta^{2}+\iota p_{0} x/
\hbar}\pm e^{-x^{2}/2 \delta^{2}-\iota p_{0} x/
\hbar}}{\sqrt{2}\pi^{1/4}\delta^{1/2}\left[
1\pm e^{-p_{0}^{2}\delta^{2}/\hbar^{2}}\right]^{1/2}}.
\end{equation}
\noindent
Here, $x_{0}$ and $p_{0}$ are real; $\delta$ can be taken to be
real for convenience. It is worth noting that superposition of the above states yields
the  single particle compass state, considered by Zurek \cite{Zurek}. 
The entangled compass state obtained here is given by,
\begin{eqnarray}
\Psi(x_{1},x_{2})&=&{\cal N}[A [\psi^{+}(x_{1}) \phi^{+}(x_{2})+\phi^{+}(x_{1})
\psi^{+}(x_{2})] + B [\psi^{-}(x_{1})
\phi^{-}(x_{2})+\phi^{-}(x_{1})\psi^{-}(x_{2})]+\nonumber \\ & &
C[\psi^{+}(x_{1}) \phi^{-}(x_{2})+\phi^{-}(x_{1})\psi^{+}(x_{2})
]+D[\psi^{-}(x_{1}) \phi^{+}(x_{2})+\phi^{+}(x_{1})\psi^{-}(x_{2})]] ,
\end{eqnarray}
where, ${\cal N}$ is the normalization constant.

In order to study the phase space structure, one computes the correlation
function:
\begin{equation}
 c(x_1,a_1,x_2,a_2)=\Psi^{\dagger}\left(
x_{1}+\frac{a}{2},x_{2}+\frac{b}{2}\right)\\
\Psi\left( x_{1}-\frac{a}{2},x_{2}-\frac{b}{2}\right).
\end{equation}
\noindent

The Wigner function,
\begin{equation}
W(x_{1}, p_{1} ; x_{2}, p_{2})
=\frac{1}{(2\pi\hbar)^{2}}{\int_{-\infty}^{\infty}}{\int_{-\infty}^{\infty}}
c(x_1,a_1,x_2,a_2)
e^{\frac{i(p_{1}a+p_{2}b)}{\hbar}}da db  .
\end{equation}
can be computed:

\begin{equation}
W(x_{1}, p_{1} ; x_{2}, p_{2})=
\frac{2\delta^{2}c|{\cal
N}|^{2}}{\pi\hbar^{2}}e^{-\frac{(x_{1}^{2}+x_{2}^{2})}{\delta^{2}}-\frac{(p_{1}^
{2}+p_{2}^{2})\delta^{2}}{\hbar^{2}}}
\sum_{i=1}^{8}\left[(W_{Di}+\sum W_{Ci})+ e^{-\frac{x_{0}^{2}}{2\delta^{2}}-\frac{p_{0}^{2}\delta^{2}}{2\hbar^{2}}}(\sum
W_{Ei})\right],
\end{equation}
where, $W_{Di}$ 's represent the diagonal components, while  $W_{Ci}$ 's and
$W_{Ei}$ 's are the off-diagonal components.

 We now carefully analyze the individual representative terms:
\begin{widetext}
\begin{eqnarray}
W_{D1}&=&|A|^{2}(e^{-\frac{x_{0}^{2}}{\delta^{2}}-\frac{p_{0}^{2}\delta^{2}}{
\hbar^{2}}}\cosh\left(\frac{2p_{0}p_{2}\delta^{2}}{\hbar^{2}}
\right) \cosh\left(\frac{2x_{0} x_{1}}{\delta^{2}} \right)+
e^{-\frac{x_{0}^{2}}{\delta^{2}}}\cosh\left(\frac{2x_{0}
x_{1}}{\delta^{2}} \right)\cos\left(
\frac{2p_{0}x_{2}}{\hbar}\right)+\nonumber \\ & &
e^{-\frac{p_{0}^{2}\delta^{2}}{\hbar^{2}}}\cos\left( \frac{2x_{0}
p_{1}}{\hbar}\right)\cosh\left(\frac{2p_{0}p_{2}\delta^{2}}{\hbar^{2}}
\right) +2\cos\left( \frac{2p_{0}x_{2}}{\hbar}\right)\cos\left(
\frac{2x_{0} p_{1}}{\hbar}\right)),
\end{eqnarray}
\end{widetext}

\begin{widetext}
\begin{eqnarray}
W_{D2}&=&|A|^{2}(
e^{-\frac{x_{0}^{2}}{\delta^{2}}-\frac{p_{0}^{2}\delta^{2}}{\hbar^{2}}}
\cosh\left(\frac{2p_{0}p_{1}\delta^{2}}{\hbar^{2}}
\right)  \cosh\left(\frac{2x_{0} x_{2}}{\delta^{2}}
\right)+e^{-\frac{x_{0}^{2}}{\delta^{2}}}\cosh\left(\frac{2x_{0}
x_{2}}{\delta^{2}} \right)\cos\left(
\frac{2p_{0}x_{1}}{\hbar}\right)+\nonumber \\ & &
e^{-\frac{p_{0}^{2}\delta^{2}}{\hbar^{2}}}\cos\left( \frac{2x_{0}
p_{2}}{\hbar}\right)\cosh\left(\frac{2p_{0}p_{1}\delta^{2}}{\hbar^{2}}
\right) +2\cos\left( \frac{2p_{0}x_{1}}{\hbar}\right)\cos\left(
\frac{2x_{0} p_{2}}{\hbar}\right)),
\end{eqnarray}
\end{widetext}

\begin{widetext}
\begin{eqnarray}
W_{C1}&=&A^{*}B(e^{-\frac{x_{0}^{2}}{\delta^{2}}-\frac{p_{0}^{2}\delta^{2}}{
\hbar^{2}}}\sinh\left(\frac{2p_{0}p_{2}\delta^{2}}{\hbar^{2}}
\right) \cosh\left(\frac{2x_{0} x_{1}}{\delta^{2}} \right)+
e^{-\frac{x_{0}^{2}}{\delta^{2}}}\sinh\left(\frac{2x_{0}
x_{1}}{\delta^{2}} \right)\cos\left(
\frac{2p_{0}x_{2}}{\hbar}\right)+\nonumber \\ & &
e^{-\frac{p_{0}^{2}\delta^{2}}{\hbar^{2}}}\sin\left( \frac{2x_{0}
p_{1}}{\hbar}\right)\cosh\left(\frac{2p_{0}p_{2}\delta^{2}}{\hbar^{2}}
\right) +2\sin\left( \frac{2p_{0}x_{2}}{\hbar}\right)\cos\left(
\frac{2x_{0} p_{1}}{\hbar}\right)),
\end{eqnarray}
\end{widetext}
and
\begin{widetext}
\begin{eqnarray}
W_{C2}&=&A^{*}B(e^{-\frac{x_{0}^{2}}{\delta^{2}}-\frac{p_{0}^{2}\delta^{2}}{
\hbar^{2}}}\sinh\left(\frac{2p_{0}p_{1}\delta^{2}}{\hbar^{2}}
\right) \cosh\left(\frac{2x_{0} x_{2}}{\delta^{2}} \right)+
e^{-\frac{x_{0}^{2}}{\delta^{2}}}\sinh\left(\frac{2x_{0}
x_{2}}{\delta^{2}} \right)\cos\left(
\frac{2p_{0}x_{2}}{\hbar}\right)+\nonumber \\ & &
e^{-\frac{p_{0}^{2}\delta^{2}}{\hbar^{2}}}\sin\left( \frac{2x_{0}
p_{2}}{\hbar}\right)\cosh\left(\frac{2p_{0}p_{1}\delta^{2}}{\hbar^{2}}
\right) +2\sin\left( \frac{2p_{0}x_{1}}{\hbar}\right)\cos\left(
\frac{2x_{0} p_{2}}{\hbar}\right)).
\end{eqnarray}
\end{widetext}

It can be seen from above that in every expression, the first three terms
containing hyperbolic functions 
are multiplied by Gaussian factors, which are bound to be small, in
the present mesoscopic context, concerned with relatively larger values of $x_0$
and $p_0$.
Thus only the last terms give a dominant contribution. 
These terms are of purely oscillating nature, which can produce  significant amount of
interference. It is to be noted that the
first two terms are similar to what had been found in Ref. \cite{panigrahi}; however the
last two terms containing $2\sin\left( \frac{2p_{0}x_{2}}{\hbar}\right)\cos\left(
\frac{2x_{0} p_{1}}{\hbar}\right)$ and $+2\sin\left( \frac{2p_{0}x_{1}}{\hbar}\right)\cos\left(
\frac{2x_{0} p_{2}}{\hbar}\right)$ are new and arise due to further interference between two groups
of terms. Explicit computation yields,
\begin{widetext}
\begin{eqnarray}
W_{E1}&=&\frac{|A|^{2}}{2} e^{\frac{i
p_{0}x_{0}}{\hbar}}(\cosh\left( (\frac{x_{0}}{\delta^{2}}-\frac{i
p_{0}}{\hbar})(x_{1}+x_{2})+(\frac{ix_{0}}{\hbar}-\frac{p_{0}\delta^{2}}{\hbar^{
2}})(p_{1}-p_{2})\right)
+\nonumber \\ & & \cosh\left( (\frac{x_{0}}{\delta^{2}}-\frac{i
p_{0}}{\hbar})(x_{1}-x_{2})+(\frac{ix_{0}}{\hbar}-\frac{p_{0}\delta^{2}}{\hbar^{
2}})(p_{1}+p_{2})\right))+\nonumber
\\ & & e^{-\frac{i p_{0}x_{0}}{\hbar}}(\cosh\left(
(\frac{x_{0}}{\delta^{2}}+\frac{i
p_{0}}{\hbar})(x_{1}+x_{2})+(\frac{ix_{0}}{\hbar}+\frac{p_{0}\delta^{2}}{\hbar^{
2}})(p_{1}-p_{2})\right)+\nonumber
\\ & & \cosh\left( (\frac{x_{0}}{\delta^{2}}+\frac{i
p_{0}}{\hbar})(x_{1}-x_{2})+(\frac{ix_{0}}{\hbar}+\frac{p_{0}\delta^{2}}{\hbar^{
2}})(p_{1}+p_{2})\right)
)+\nonumber \\ & &
2(\cos\left(p_{0}(\frac{(x_{1}-x_{2})}{\hbar}-\frac{i
(p_{1}+p_{2})\delta^{2}}{\hbar^{2}}) \right)\cosh\left(
x_{0}(\frac{(x_{1}+x_{2})}{\delta^{2}}+\frac{i(p_{1}-p_{2})}{\hbar}
)\right)+\nonumber \\ & &
\cos\left(p_{0}(\frac{(x_{1}+x_{2})}{\hbar}-\frac{i
(p_{1}-p_{2})\delta^{2}}{\hbar^{2}}) \right)\cosh\left(
x_{0}(\frac{(x_{1}-x_{2})}{\delta^{2}}+\frac{i(p_{1}+p_{2})}{\hbar}
)\right)))
\end{eqnarray}
\end{widetext}

where the purely oscillating terms are absent.
Combining the oscillatory terms from all the expressions one gets in the mesoscopic limit:
\begin{eqnarray}
 W \approx & & A_{1}\cos\left( \frac{2p_{0}x_{2}}{\hbar}\right)\cos\left(
\frac{2x_{0} p_{1}}{\hbar}\right) + A_{2}\cos\left(
\frac{2p_{0}x_{1}}{\hbar}\right)\cos\left(
      \frac{2x_{0} p_{2}}{\hbar}\right)+\nonumber \\& & A_{3}\sin\left(
\frac{2p_{0}x_{2}}{\hbar}\right)\cos\left(
\frac{2x_{0} p_{1}}{\hbar}\right) + A_{4}\sin\left(
\frac{2p_{0}x_{1}}{\hbar}\right)\cos\left(
      \frac{2x_{0} p_{2}}{\hbar}\right),
\end{eqnarray}
Where $A_1,A_2,A_3$ and $A_{4}$ are related to A, B, C and D. The precise relationship is not important for the sub-Planck structure.
It is worth noting that in the $x_{1}p_{1}$ plane, where both $x_{2}$ and $p_{2}$ are kept constant we find,
\begin{equation}
W \approx B_{1}\cos\left(\frac{2x_{0} p_{1}}{\hbar}\right) + B_{2}\cos\left(
\frac{2p_{0}x_{1}}{\hbar}\right) + B_{3}\sin\left(
\frac{2p_{0}x_{1}}{\hbar}\right).
\end{equation}
The distance between two zeros in $x_1$ direction is 
$\pm\frac{\pi\hbar}{4p_{0}}$,
while it is  $\pm\frac{\pi\hbar}{4x_{0}}$ in the $p_1$-direction. This gives the
area of the fundamental
tile $a=\frac{(2\pi\hbar)^{2}}{4x_{0} p_{0}}$ in $x_1p_1$ plane of particle one.
Similarly one can find zeros in $x_2$
and $p_2$ directions and obtain the same value of the fundamental area.  
It is clear that visibility of the checkerboard type interference patterns
originates from two sources. It arises due to the entanglement of
$|\alpha^{+}\rangle, |\alpha^{-}\rangle, |\iota\alpha^{+}\rangle$ and
$|\iota\alpha^{-}\rangle$, as well as from the interference of these entangled
states. As has been pointed out by Toscano \it et al. \normalfont in \cite{Toscano}, the sensitivity of this
state to external perturbations can be checked by the
following scheme.
Let $D_1(\alpha)$ and  $D_2(\beta)$ denote two displacement
operators causing the displacement of particle states one and two, by amount
$\alpha$
and $\beta$ respectively, to create a perturbed state
$\vert\psi_{per}\rangle=D_1(\alpha)D_2(\beta)
\vert\psi_c\rangle$. In the case, where there are equal shifts for both the
particles i.e., $\alpha=\beta=is\frac{x_0}{\vert x_0\vert} $, the overlap
function takes the form ,
\begin{equation}
\vert
\langle\psi_c\vert\psi_{per}\rangle\vert^2\propto\left(1+cos\left(4x_0s + \theta
\right)\right).
\end{equation}
\noindent 
For a fixed $\theta$ and $x_0$, the vanishing of the right hand side of Eq.(34) occurs at $s\sim (\frac{\pi}{(4x_0)} - \theta) $. In the case of the state considered in Ref.\cite{panigrahi}, $\theta$ was 0. In general, $\theta$ can have a non-zero value arising due to the contribution of off-diagonal elements to the zeroes of the Wigner function. Thus, it enables one to carry out Heisenberg-limited
measurements. It is clear that this is due to the presence of sub-Planck structure in phase space.

\section{Conclusion}
In conclusion, a scheme has been proposed to generate entangled bipartite states, 
with compass type sub-Planck sensitivity, through the trapping of two identical bosonic atoms in a
double well potential in the Lamb-Dicke regime. The Hamiltonian selects out
macroscopic states, which couple with the internal degrees of freedom, when the
system is in its ground state. Tunneling between the two wells select out the symmetric state as the ground state of the system. An appropriate measurement then
generates the desired state with Heisenberg-limited sensitivity. The fact that
this generic Hamiltonian has found applications in many physical systems, makes
the present proposal experimentally feasible. In a recent study, quantum
interference structures
have been observed in trapped ion dynamics, beyond the Lamb-Dicke approximation
\cite{Wang}. These systems may also show sub-Planck
structures and hence need careful study. Our analysis has been done with vanishing detuning parameter. For a small value of the same, we expect that
the state produced would still exhibit sub-Planck structures. However, this
requires detailed numerical investigation. We propose to study the effect of decoherence on this system. We hope that our proposal will be
realized in optical traps and cold atoms.

\end{document}